\documentclass[12pt,preprint]{aastex}
\usepackage{pslatex}
\begin{document}

\title{Particle acceleration by slow modes in strong compressible MHD turbulence,
with application to solar flares.}
\author{Benjamin D. G. Chandran }
\email{ benjamin-chandran@uiowa.edu} 
 \affil{Department of Physics \& Astronomy, University of Iowa}

\begin{abstract}
Energetic particles that undergo strong pitch-angle scattering and
diffuse through a plasma containing strong compressible MHD turbulence
undergo diffusion in momentum space with diffusion
coefficient~$D_p$. If the rms turbulent velocity is of order the
Alfv\'en speed~$v_{\rm A}$, the contribution to~$D_p$ from slow-mode
eddies is~$\simeq (2p^2 v_{\rm A}/9 l)[\ln (l v_{\rm A}/D_\parallel) +
2\gamma - 3]$, where $l$ is the outer scale of the turbulence,
$\gamma\simeq 0.577$ is Euler's constant, and~$D_\parallel$ is the
spatial diffusion coefficient of energetic particles, which is assumed
to satisfy~$D_\parallel \ll l v_{\rm A}$.  The energy spectrum of
accelerated particles is derived for this value of~$D_p$, taking into
account Coulomb losses and particle escape from the acceleration
region with an energy-independent escape time.

Slow modes in the $D_\parallel \ll l v_{\rm A}$-limit are an unlikely
explanation for electron acceleration in solar flares to energies of
10-100~keV, because for solar-flare conditions the predicted
acceleration times are too long and the predicted energy spectra are
too hard.  The acceleration mechanism discussed in this paper could in principle
explain the relatively hard spectra of gyrosynchrotron-emitting
electrons in the~100-5000~keV range, but only if $D_\parallel \ll l
v_{\rm A}$ for such particles.

\end{abstract}
\maketitle

\section{Introduction}
\label{sec:intro} 

The first part of this paper treats stochastic particle acceleration
by slow modes in compressible magnetohydrodynamic (MHD)
turbulence assuming efficient pitch-angle scattering. 
The second part presents the energy spectra of
accelerated particles taking into account Coulomb losses and
assuming that particles escape the acceleration region on
an energy-independent time scale. The third part of the paper
shows that slow-modes in the efficient-pitch-angle-scattering
limit are not a viable explanation for electron acceleration
in solar flares in the 10-100~keV energy range, although 
in principle they could explain the electron spectra at 100-5000~keV
inferred from microwave gyrosynchrotron radiation.

Throughout the paper it is assumed that energetic particles propagate through a magnetized
plasma containing both ``large-scale'' turbulent compressible motions
that vary over distances much larger than the energetic-particle
gyroradius~$\rho_g$ and ``small-scale'' waves that vary over
distances~$\sim \rho_g$. The small-scale waves scatter particles,
causing diffusion in physical space with mean free path~$\lambda_{\rm
mfp}$ as well as diffusion in momentum space. The large-scale motions
also cause spatial and momentum diffusion.  Only those large-scale
plasma motions on scales $\gg \lambda_{\rm mfp}$ are considered.
Particle acceleration by velocities on scales $\gg \lambda_{\rm mfp}$
has been considered previously by a number of authors [e.g., Kulsrud
\& Ferrari (1971) Bykov \& Toptygin (1982, 1990, 1993), Ptuskin
(1988), Dolginov \& Silantev 1990, Katz \& Stehlik (1991),
Chandran \& Maron (2003)]. In the present paper, the large-scale
velocities are taken to be slow-mode eddies in strong anisotropic
compressible MHD turbulence. Such eddies are elongated along the local
magnetic field, and involve velocities directed primarily along the
local magnetic field (Lithwick \& Goldreich 2001, Cho \& Lazarian
2002).  Part of the motivation for considering slow modes is that
under certain conditions they can contribute significantly to momentum
diffusion.  An additional motivation is for completeness; by
comparison the effects of fast modes and Alfv\'en modes are relatively
well understood (e.g., Jokipii 1966, Kulsrud \& Pearce 1969,
Berezinskii et~al 1990,
Miller et~al 1997, Schlickeiser \& Miller 1998,
Chandran 2000, Schlickeiser 2002, Yan \& Lazarian 2002).

If the small-scale waves are Alfv\'en waves, 
the transport equation describing the energetic-particle distribution
function~$f$ is (Skilling 1975)
\begin{equation}
\frac{\partial f}{\partial t} + \left[ \frac{\partial}{\partial p^3}
\left(p^3 {\bf w }\right)\right] \cdot \nabla f 
= \nabla \cdot \left( D_\parallel \hat{\bf b} \hat{\bf b} \cdot \nabla f\right)
+ (\nabla \cdot {\bf w})
\frac{p}{3}\, \frac{\partial f}{\partial p}
+ \frac{\partial}{\partial p^3}\left[9 p^4 D_{\rm pp}^{\rm (s)} \frac{\partial f}{
\partial p^3}\right],
\label{eq:ski1} 
\end{equation} 
where $p$ is momentum, $\hat{\bf b}$ is the magnetic-field unit vector,
\begin{equation}
{\bf w} = {\bf u} + \left\langle
\frac{3}{2} (1-\mu^2) \frac{\nu_+ - \nu_-}{\nu_+ + \nu_-}\right\rangle v_{\rm A} \hat{\bf b}
\label{eq:defw} 
\end{equation} 
is an effective ``wave-frame'' velocity 
at which the cosmic rays are advected, ${\bf u}$ is the
large-scale plasma velocity, $v_{\rm A}$ is the Alfv\'en speed,
$\nu_+$ and $\nu_-$ are, respectively, the pitch-angle scattering rates
associated with waves traveling parallel to and anti-parallel to the
magnetic field, $\mu$ is the pitch-angle cosine,
$\langle \dots \rangle$ indicates an average over~$\mu$,
\begin{equation}
D_\parallel = v^2 \left\langle\frac{1-\mu^2}{2(\nu_+ + \nu_-)}\right \rangle
\label{eq:defdp} 
\end{equation} 
is the diffusion coefficient for particle motion along the magnetic field,
\begin{equation}
D_{pp}^{\rm (s)} = 4 \gamma_L^2 m^2 v_{\rm A}^2
\left\langle\left( \frac{1-\mu^2}{2} \right)\frac{\nu_+ \nu_-}{\nu_+ + \nu_-}\right\rangle
\label{eq:defdpps} 
\end{equation} 
is the momentum diffusion coefficient associated with the small-scale
waves, and~$\gamma_L$ is the relativistic Lorentz factor of an
energetic particle.  Equations~(\ref{eq:defw}) and (\ref{eq:defdpps})
show that either~${\bf w} \neq {\bf u}$ or $D_{ pp}^{\rm s} \neq
0$ (Schlickeiser 2002). Nevertheless, for simplicity the momentum diffusion associated
with small-scale waves is ignored ($D_{pp}^{\rm s} \rightarrow 0$),
and at the same time, inconsistently, it is assumed that ${\bf w} =
{\bf u}$.  This approximation has been standard in studies of particle
acceleration by velocities on scales~$\gg \lambda_{\rm mfp}$ [Bykov \&
Toptygin (1982, 1990, 1993), Ptuskin (1988), Dolginov \& Silantev
1990, Katz \& Stehlik (1991)]; it reduces
equation~(\ref{eq:ski1})  to
\begin{equation}
\frac{\partial f}{\partial t} + 
{\bf u } \cdot \nabla f 
= \nabla \cdot \left( D_\parallel \hat{\bf b} \hat{\bf b} \cdot \nabla f\right)
+ (\nabla \cdot {\bf u})
\frac{p}{3}\, \frac{\partial f}{\partial p}.
\label{eq:ski2} 
\end{equation} 

Equation~(\ref{eq:ski2}) is used to calculate the momentum diffusion
coefficient arising from the large-scale turbulence.  It seems
plausible that a nonzero $D_{pp}^{\rm s}$ would approximately add to
the momentum diffusion coefficient associated with the large-scale
turbulence calculated from equation~(\ref{eq:ski2}), but future work is needed to
verify this. The spatial diffusion caused by the
large-scale turbulence is neglected.  It is assumed that the rms
large-scale turbulent velocity is comparable to~$v_{\rm A}$ and that
the ratio~$\beta$ of thermal to magnetic pressure is~$\lesssim 1$. The
power spectrum of the slow modes is assumed to follow the theory of
Lithwick \& Goldreich (2001), which has received support from direct
numerical simulations (Cho \& Lazarian 2002).  It is also assumed that
$v_{\rm A} d_{\rm min} \ll D_\parallel \ll v_{\rm A} l$, where
$d_{\rm min}$ is the length along the magnetic field of the smallest
slow-mode eddies under consideration ($d_{\rm min}$ is either the
length of the eddies at the dissipation scale or several times~$\lambda_{\rm mfp}$,
whichever is larger), and~$l$ is the length of the largest
(outer-scale) eddies. Since the time for an eddy of
length~$\lambda_\parallel$ (measured along the magnetic field)
to be randomized
is~$\lambda_\parallel/v_{\rm A}$ (Lithwick \& Goldreich 2001), the
inequality $v_{\rm A} d_{\rm min} \ll D_\parallel \ll v_{\rm A} l$
implies that the largest eddies are randomized before
a particle diffuses through them, and that a particle diffuses through
eddies of length~$d_{\rm min}$ before such eddies are randomized.  The
assumption $D_\parallel \gg v_{\rm A} d_{\rm min}$ requires that the
particle speed~$v$ satisfy $v\gg v_{\rm A}$ since $D_\parallel \sim
v\lambda_{\rm mfp}$ and $d_{\rm min} > \lambda_{\rm mfp}$.

The procedure used to calculate the momentum diffusion
coefficient~$D_p$ arising from the large-scale velocities is to
average over the slow-mode fluctuations, neglect spatial variations
in the  averaged distribution function~$f_0$, and approximate
the eddies on scales comparable to~$l$ as a uniform background
field~${\bf B} _0$. The eddies on scales~$\ll l$ have velocities~$\ll
v_{\rm A}$ and magnetic perturbations~$\ll B_0$ and can be treated
using quasilinear theory. It is found that
\begin{equation}
\frac{\partial f_0}{\partial t} = \frac{1}{p^2} \frac{\partial
}{\partial p} \left( p^2 D_p \frac{\partial f_0}{\partial p}\right),
\label{eq:fp0} 
\end{equation} 
with
\begin{equation} 
D_p \simeq \frac{2 p^2 v_{\rm A}}{9  l}\; \left[\ln 
\left(\frac{l v_{\rm A}}{D_\parallel}\right) + 2\gamma - 3 \right],
\mbox{ \hspace{0.3cm} for $v_{\rm A} d_{\rm min} \ll D_\parallel \ll v_{\rm A} l$,}
\label{eq:dpintro} 
\end{equation} 
where $\gamma \simeq 0.577$ is Euler's constant.  The corresponding
acceleration time scale,~$p^2/D_p$, is of order the large-eddy
turnover time $l/v_{\rm A}$, and depends only weakly through the
$\ln(l v_{\rm A}/D_\parallel)$ term on~$p$ and particle species.  An
approximate version of equation~(\ref{eq:dpintro}) is obtained using
phenomenological arguments in section~\ref{sec:phen}.

Section~\ref{sec:spectra} presents the time-dependent spectrum of
accelerated particles assuming an ad-hoc model of particle escape from
the acceleration region with an energy-independent escape time.
Section~\ref{sec:stst} gives the steady-state spectrum for electron
acceleration taking into account Coulomb losses, again assuming an
energy-independent escape time.  Section~\ref{sec:sf} compares the
predicted spectra and acceleration times to observations of electrons
in solar flares.  The results of the paper are summarized in
section~\ref{sec:conc}.  Much of the notation used in the paper is
defined in table~\ref{tab:t1}.

\begin{table*}[h]
\begin{tabular}{lc}
\hline
\hline 
&
\vspace{-0.25cm} 
\\
Notation & Meaning \vspace{0.1cm} \\
\hline  
\vspace{-0.2cm} \\
$\lambda_{\rm mfp}$ & scattering mean free path of energetic particles \\
$\lambda_{\rm thermal}$ & Coulomb mean free path of thermal protons \\
$\lambda_\parallel$ & length of a turbulent eddy measured along the magnetic field \\
$\lambda_\perp$ &  width of a turbulent eddy measured across the magnetic field \\
$r_g$ & energetic-particle gyroradius \\
$f$ & energetic-particle distribution function \\
$N(E)$ & number of accelerated particles per unit energy \\
${\bf B} $ & magnetic field \\
$d_\parallel$ & length along magnetic field of eddies at the dissipation scale \\
$d_{\rm min}$ &  $d_\parallel$ or several times~$\lambda_{\rm mfp}$, whichever is larger \\
$l$ & stirring scale (outer scale) of turbulence \\
$v_{\rm A} $ & Alfv\'en speed \\
$c_{\rm s}$ & sound speed \\
$\beta$ & ratio of thermal to magnetic pressure \\
${\bf u} $ & turbulent velocity \\
$v$ & energetic-particle velocity \\
$D_\parallel$ & diffusion coefficient for motion along magnetic field \\
$D_p$ & momentum diffusion coefficient \\
$\tau_{\rm esc}$ & escape time scale \\
$\tau_{\rm acc}$ & acceleration time scale \\
$\tau$ & $t/\tau_{\rm acc}$ \\
$\chi$ & $\tau_{\rm acc}/\tau_{\rm esc}$ \\
$\gamma$ & Euler's constant, $\sim 0.577$ \vspace{0.2cm} \\
\hline
\hline
\end{tabular}
\caption{Definitions.
\label{tab:t1} }
\end{table*}

\section{Stochastic particle acceleration by slow-mode eddies in
strong compressible MHD turbulence}
\label{sec:sa} 

This section reviews properties of small-amplitude slow magnetosonic
waves and compressible MHD turbulence and presents phenomenological
and analytic derivations of the contribution to~$D_p$ from slow modes
in compressible MHD turbulence.

\subsection{Properties of small-amplitude slow magnetosonic waves}
\label{sec:slow} 

It is instructive to consider the properties of a small-amplitude slow
magnetosonic wave propagating in a uniform background magnetic
field~${\bf B} _0 =B_0 \hat{\bf z}$ with frequency~$\omega$, wave
vector~${\bf k} $, and $k_z\ll k$.  Such a wave varies much more
rapidly across the magnetic field than along the magnetic field, like
slow modes in compressible MHD turbulence (see
section~\ref{sec:MHD}).  The wave frequency, polarization, and
velocity divergence when $k_z \ll k$ are given by (Lithwick \& Goldreich
2001)
\begin{equation}
\omega = \frac{c_{\rm s}v_{\rm A} |k_z|}{\sqrt{c_{\rm s}^2 + v_{\rm A}^2}} \left[
1 + {\cal O}\left( \frac{k_z}{k}\right)^2\right],
\label{eq:omega} 
\end{equation} 
\begin{equation}
\frac{u_x}{u_z} \simeq  -\frac{c_{\rm s}^2}{c_{\rm s}^2 + v_{\rm A}^2 } \;
\frac{k_z}{k}, 
\label{eq:ux} 
\end{equation} 
\begin{equation}
u_y = 0,
\label{eq:uy} 
\end{equation} 
and 
\begin{equation}
{\bf k} \cdot {\bf u} = \frac{v_{\rm A} ^2}{c_{\rm s}^2 + v_{\rm A}^2}
 \left[
1 + {\cal O}\left( \frac{k_z}{k}\right)^2\right]
k_z u_z,
\label{eq:divv} 
\end{equation} 
where ${\bf u}$ is the fluctuating velocity,
${\bf k}$ is taken to lie in the $xz$-plane, and $c_{\rm s}$ is the
sound speed. 
Equations~(\ref{eq:ux}) and (\ref{eq:uy}) imply that the slow-wave
velocity is approximately aligned with the background magnetic field.
The total pressure perturbation (magnetic plus thermal) 
vanishes to second order in $k_z/k$ (Lithwick \& Goldreich 2001).
An important
property of slow waves with $k_z \ll k$ when
$\beta \lesssim 1$ (and thus $c_{\rm s} \lesssim v_{\rm A}$)
follows from equations~(\ref{eq:ux}) and (\ref{eq:divv}):
${\bf k} \cdot
{\bf u} \sim k_z |{\bf u}|$.

\subsection{Slow-mode eddies in strong anisotropic MHD turbulence}
\label{sec:MHD}

The anisotropy of magnetohydrodynamic (MHD) turbulence has been
studied by many authors.\footnote{ See, e.g., Montgomery \&
Turner 1981, Shebalin et~al.\ 1983, Higdon 1984, Higdon 1986,
Oughton et~al.\ 1994, Sridhar \& Goldreich 1994, Goldreich \&
Sridhar 1995, Montgomery \& Matthaeus 1995, Ghosh \& Goldstein
1997, Goldreich \& Sridhar 1997, Matthaeus et~al.\ 1998, Spangler
1999, Bhattacharjee \& Ng 2000, Cho \& Vishniac 2000, Maron \&
Goldreich 2001, Milano et~al 2001, Lithwick \& Goldreich 2001, Cho \& Lazarian 2002.} In
this paper, it is assumed that the ambient plasma is stirred at a
scale~$l$, and that the rms turbulent velocity is comparable
to~$v_{\rm A}$.  In this case, the rms magnetic fluctuation at
scale~$l$, denoted $B_l$, is comparable to any mean magnetic
field in the system.  The stirring excites an inertial range of
turbulent fluctuations extending from the large scale~$l$ to
a much smaller dissipation scale. Within any
box of dimension $\ll l$, the fluctuations can be decomposed into
the three MHD wave polarizations with respect to the average
magnetic field direction within the box, ${\bf B} _{\rm
local}$. For example, when $\beta \ll 1$, the velocity fluctuations
aligned with ${\bf B}_{\rm local}$ are associated with slow waves.
The slow-wave fluctuations within the box can be thought of as
a collection of
nested eddies, where an eddy is simply a volume of some specified
width~$\lambda_\perp$ measured across the magnetic field and
length~$\lambda_\parallel$ measured along the magnetic field,
for which the velocity variation across the width of the eddy is
comparable to the velocity variation along the length of the eddy.
For values of~$\lambda_\perp$ in the inertial range, 
slow-mode eddies are elongated  along ${\bf B}
_{\rm local}$, with
\begin{equation}
\lambda_\parallel \sim \lambda_\perp^{2/3} l^{1/3},
\label{eq:lperp} 
\end{equation} 
and the rms velocity variation across a slow-mode eddy is
\begin{equation}
u_{\lambda_\perp} \sim v_{\rm A} \left(\frac{\lambda_\perp}{l}\right)^{1/3}
\label{eq:u0} 
\end{equation} 
(Lithwick \&
Goldreich 2001, Cho \& Lazarian 2002).
Equation~(\ref{eq:lperp}) means that the velocity varies
more rapidly across~${\bf B}_{\rm local}$ than along~${\bf B}_{\rm local}$.
Suppose the vector separation between two points is~${\bf r}$, 
with~$r$ in the inertial range of the turbulence, and let
the rms velocity difference between the two points be~$\delta {\bf u}$.
If ${\bf r}\perp {\bf B}_{\rm local}$, equation~(\ref{eq:u0}) implies that~$\delta u 
\sim v_{\rm A} (r/l)^{1/3}$; if ${\bf r} \parallel {\bf B}_{\rm local}$, 
equations~(\ref{eq:lperp}) and (\ref{eq:u0}) imply that $
\delta u \sim v_{\rm A} (r/l)^{1/2}$.
Slow-mode eddies are randomized in a time~$\sim
\lambda_\parallel/v_{\rm A}$ due to the mixing of slow modes
by Alfv\'en-mode eddies~(Lithwich \& Goldreich 2001).

The dissipation scale in the directions perpendicular to~${\bf B}_{\rm local}$,
denoted $d_\perp$, and the corresponding parallel scale~$d_\parallel =d_\perp^{2/3} l^{1/3}$,
are given by (Lithwick \& Goldreich 2001)
\begin{equation} 
d_\perp  \sim  r_{\rm g,\, thermal} \mbox{ \hspace{0.3cm} for $\beta \ll 1$, and} \\
\label{eq:diss1} 
\end{equation} 
\begin{equation} 
d_\parallel  \sim \lambda_{\rm thermal} \mbox{ \hspace{0.3cm} for $\beta \gtrsim 1$},
\label{eq:diss2} 
\end{equation} 
where $\lambda_{\rm thermal}$ is the collisional mean free path of
thermal ions, and ion-neutral friction and radiative cooling are
ignored.  Equation~(\ref{eq:diss1}) is surprising in that linear
slow magnetosonic waves are strongly damped in low-$\beta$
plasmas on scales smaller than the collisional mean free path:
i.e., when $\beta\ll 1$, the damping time of slow modes with
$\lambda_\parallel < \lambda_{\rm thermal}$ is~$\sim
\lambda_\parallel/c_{\rm s}$, comparable to the linear wave
period.  The reason that turbulent slow-mode eddies can persist
on parallel scales $< \lambda_{\rm thermal}$ when $\beta\sim
c_{\rm s}^2/v_{\rm A} ^2 \ll 1$ is that the cascade of slow mode
eddies is controlled by the Alfv\'en modes, and the cascade
time~$\lambda_\parallel/v_{\rm A}$ is much less than the linear
slow-wave damping time~(Lithwick \& Goldreich 2001).

\subsection{Phenomenological estimate of $D_p$}
\label{sec:phen} 

Since only that part of the turbulence with $\lambda_\parallel
\gg \lambda_{\rm mfp}$ is considered,  and since the momentum diffusion
associated with small-scale waves is ignored, the time derivative of
a particle's momentum induced by the large-scale velocity~${\bf u}$
is given by (Ptuskin 1988)
\begin{equation}
\frac{dp}{dt} = - \frac{\nabla \cdot u}{3} \;\; p.
\label{eq:dpdt} 
\end{equation} 
For $\beta \lesssim 1$, the rms velocity divergence of slow-mode eddies of
width~$\lambda_\perp$ satisfies (Lithwick \& Goldreich 2001)
\begin{equation}
\langle |\nabla \cdot {\bf u}|_{\lambda_\perp}^2 \rangle^{1/2} \sim
\frac{u_{\lambda_\perp}}{\lambda_\parallel }.
\label{eq:divu0} 
\end{equation}
The rms contribution to $dp/dt$ from eddies of length $\lambda_\parallel$ is
thus
\begin{equation}
\left(\frac{dp}{dt} \right)_{\lambda_\parallel}
\sim \frac{p v_{\rm A}}{3\sqrt{ \lambda_\parallel l}}.
\label{eq:dpdt2} 
\end{equation} 
If a particle interacts coherently with an eddy of length
$\lambda_\parallel$ for a time $\triangle t$, it incurs a random
momentum increment of rms magnitude $\triangle p \sim (dp/dt)
\triangle t$. The contribution to~$D_p$ from eddies of
size~$\lambda_\parallel$ is $ \sim (\triangle p)^2/\triangle t$.
When $d_{\rm min} v_{\rm A} \ll D_\parallel \ll l v_{\rm A} $,
particles are confined within eddies with $\lambda_\parallel >
D_\parallel/v_{\rm A}$ throughout the time~$\lambda_\parallel/v_{\rm
A}$ required for the eddies to be randomized in the turbulent
flow. For such eddies, $\triangle t \sim \lambda_\parallel/v_{\rm
A}$. Each eddy size between~$D_\parallel/v_{\rm A}$ and~$l$ makes
a contribution to~$D_p$ of~$\sim p^2 v_A/(9l)$. The contribution
to~$D_p$ from all such eddies is
\begin{equation}
D_p \sim \frac{p^2 v_{\rm A}}{9 l} \;\ln\left(\frac{l v_{\rm A} }{D_\parallel}\right)
\mbox{ \hspace{0.3cm} for $d_{\rm min} \ll (D_\parallel/v_{\rm A}) \ll l$,}
\label{eq:dp8a} 
\end{equation} 
in approximate agreement with equation~(\ref{eq:dp4a}) below.
The contribution to~$D_p$ from eddies with $\lambda_\parallel <
D_\parallel/v_{\rm A}$ can be neglected.  
\footnote{If slow-mode eddies were oscillatory, as opposed to repeatedly
randomized, then eddies larger than~$D_\parallel/v_{\rm A}$ would
contribute less to~$D_p$ than eddies of length~$D_\parallel/v_{\rm
A}$, and the logarithmic term would disappear from
equation~(\ref{eq:dp8a}). However, although $\nabla \cdot {\bf u}$ is
oscillatory within any infinitesimal fluid element (since the density
does not change secularly in time), the average value of $\nabla\cdot
{\bf u}$ over the volume of a slow-mode eddy is randomized in the
time~$\lambda_\parallel/v_{\rm A}$ since the eddy is completely
shredded and mixed with its neighbors in a time~$\sim
\lambda_\parallel/v_{\rm A}$ by Alfv\'en modes.  In addition, an
energetic particle moves from one fluid element within an eddy of
length~$\lambda_\parallel$ to a new fluid element that is chosen
randomly from the surrounding eddy-volume of
width~$\lambda_\perp$ and length~$\lambda_\parallel$ in a
time~$\lambda_\parallel/v_{\rm A}$ due to cross-field
diffusion and the turbulent mixing of fluid elements by
Alfv\'en~modes.}

\subsection{Derivation of~$D_p$ in the quasilinear approximation.}
\label{sec:an} 

Particle acceleration by slow-mode eddies on scales $\ll l$ is now
treated analytically for $\beta \lesssim 1$.  A scale $l^\prime$ is
introduced with $\lambda_{\rm mfp} \ll l^\prime \ll l$, and an
approximation is made in which eddies with $l^\prime < \lambda_\perp <
l$ are replaced with a uniform magnetic field~$B_l \hat{\bf z}$, so
that ${\bf B}_{\rm local}$ is everywhere approximately along~$\hat{\bf
z}$.  In terms of the Fourier transform of the velocity~${\bf u}({\bf
k},t)$, and in accord with equations~(\ref{eq:lperp})
and~(\ref{eq:u0}), the spectrum of the (homogeneous) turbulence is
taken to be~(Goldreich \& Sridhar 1995, Lithwick \& Goldreich 2001,
Cho \& Lazarian 2002)
\begin{equation}
\langle {\bf u} ({\bf k}_1, t_1) \cdot {\bf u} ({\bf k}_2, t_2)\rangle
= P({\bf k}_1) e^{-\gamma_k|t_1 - t_2|} \delta ({\bf k}_1 + {\bf k}_2),
\label{eq:c1} 
\end{equation} 
with
\begin{equation} 
P({\bf k}) = \frac{v_{\rm A}^2}{6\pi} \; l^{-1/3} k_\perp^{-10/3} g\left(\frac{
k_z \,l^{1/3}}{k_\perp^{2/3}}\right)
\label{eq:P} 
\end{equation} 
for $(l^\prime)^{-1} < k_\perp < d_\perp^{-1}$ [$P({\bf k})= 0$ otherwise],
\begin{equation}
g(x) = e^{-|x|}
\label{eq:g} ,
\end{equation} 
and 
\begin{equation}
\gamma_k  = k_\perp^{2/3} l^{-1/3} v_{\rm A} \sim \frac{v_{\rm A}}{\lambda_\parallel},
\label{eq:gammak} 
\end{equation} 
where $k_\perp = \sqrt{k_x^2 + k_y^2}$. Given equation~(\ref{eq:divu0}),
it is assumed that
\[
\langle {\bf k}_1 \cdot {\bf u}({\bf k}_1, t_1)\;\;
{\bf k}_2 \cdot {\bf u}({\bf k}_2, t_2) \rangle
\]
\begin{equation}
= - k_{1z}^2 \; P({\bf k}_1) e^{-\gamma_{k1} |t_1-t_2|} \delta({\bf k}_1
+ {\bf k}_2).
\label{eq:comp} 
\end{equation} 
For ${\bf B} _{\rm local} \propto \hat{\bf z}$, equation~(\ref{eq:ski2}) 
becomes
\begin{equation}
\frac{\partial f}{\partial t} + {\bf u } \cdot \nabla f 
=  D_\parallel \frac{\partial ^2 f}{\partial z^2} + (\nabla \cdot {\bf u})
\frac{p}{3}\, \frac{\partial f}{\partial p}.
\label{eq:crt} 
\end{equation} 
The distribution function is written as the sum of two
parts,
\begin{equation}
f = f_0 + f_1,
\label{eq:tp} 
\end{equation} 
where $f_0 = \langle f \rangle$, $f_1 = f- \langle f \rangle$, and
angled brackets denote an average over an ensemble of realizations of
the turbulence. For simplicity, it is assumed that
$\langle u \rangle = 0$ and $\nabla f_0 = 0.$ The ensemble average of
equation~(\ref{eq:crt}) is then
\begin{equation}
\frac{\partial f_0}{\partial t} = - \langle {\bf u} \cdot \nabla f_1\rangle +
\left
\langle (\nabla \cdot {\bf u} ) \frac{p}{3}\,\frac{\partial f_1}{\partial p}\right\rangle.
\label{eq:f0} 
\end{equation}
Subtracting equation~(\ref{eq:f0}) from 
equation~(\ref{eq:crt}) yields
\begin{equation}
\frac{\partial f_1}{\partial t} - D_\parallel \frac{\partial ^2 f_1}
{\partial z^2} = \frac{(\nabla\cdot {\bf u}) p}{3}
\, \frac{\partial f_0}{\partial p},
\label{eq:f1} 
\end{equation} 
in which products of fluctuating quantities have been neglected.  Such
products are small compared to $|\partial f_1/\partial t|$ since the
slow-mode velocity is almost along $z$, $u \ll v_{\rm A}$ for eddies
on scales~$\ll l$, and
$|\partial f_1/\partial t| \sim v_{\rm A} |\partial f_1/\partial
z|$. Upon solving for~$f_1$ and substituting into equation~(\ref{eq:f0}),
one recovers equation~(\ref{eq:fp0})  with 
\begin{equation}
D_p = \frac{p^2}{9}\,
\int d^3 k  \int_0^{\infty} d\tau \;\;k_z^2 P({\bf k}) e^{-(\gamma_k + k_z^2 D_\parallel)\tau}.
\label{eq:dpa} 
\end{equation} 
When $d_{\rm min}
\ll (D_\parallel/v_{\rm A}) \ll l_\parallel^\prime$,
\begin{equation} 
D_p = \frac{2 p^2 v_{\rm A}}{9  l}\; \left[\ln 
\left(\frac{l_\parallel^\prime v_{\rm A}}{D_\parallel}\right) + 2\gamma - 3 \right],
\label{eq:dp4a} 
\end{equation} 
where $\gamma \simeq 0.577$ is Euler's constant.  An estimate of the
contribution to~$D_p$ from all eddies, including those with
$l_\parallel^\prime < \lambda_\parallel < l$, is obtained by taking
$l_\parallel^\prime \rightarrow l$ in equation~(\ref{eq:dp4a}).

\section{Time-dependent energy spectrum of accelerated particles
assuming an energy-independent escape time and no Coulomb losses}
\label{sec:spectra}

In this section, the energy spectrum of accelerated particles is
presented for~$p>p_0$, where $p_0$ corresponds to a particle velocity
of a few times~$v_{\rm A}$. Coulomb collisions are neglected, and
particle escape from the acceleration region is modeled in an ad hoc
fashion by adding a loss term to the right-hand side of
equation~(\ref{eq:fp0}),
\begin{equation}
\frac{\partial f}{\partial t} = \frac{1}{p^2} \frac{\partial
}{\partial p} \left( p^2 D_p \frac{\partial f}{\partial p}\right)
- \frac{f}{\tau_{\rm esc}},
\label{eq:fp0a} 
\end{equation} 
where the escape time $\tau_{\rm esc}$ is assumed to be
momentum-independent.  In equation~(\ref{eq:fp0a}) and throughout this
section, the subscript on~$f_0$ will be dropped, with the
understanding that $f$ now refers to the distribution function
obtained after averaging over the turbulent fluctuations.
Equation~(\ref{eq:fp0a}) is solved for $p\geq p_0$ and $t\geq 0$
subject to the boundary conditions
\begin{eqnarray} 
f(p=p_0, \;t\geq 0) & = & A \mbox{ \hspace{0.3cm} with $\;A\;$ constant,}
\label{eq:bc1} \\
f(p>p_0, \;t=0) & = & 0, \mbox{ \hspace{0.3cm} and} \label{eq:bc2} \\
f(p\rightarrow \infty, \;t\geq 0) & \rightarrow & 0 \label{eq:bc3} .
\end{eqnarray} 
Equation~(\ref{eq:bc1}) would apply, at least approximately, if
Coulomb collisions kept the distribution
at low energies fixed and only a small minority of the particles
were accelerated. It is assumed that
\begin{equation}
D_p = \frac{p^2}{\tau_{\rm acc}},
\label{eq:dpprop} 
\end{equation} 
with~$\tau_{\rm acc}$ independent of~$p$.
From equation~(\ref{eq:dp4a}) it can be seen that $\tau_{\rm acc}$ for
acceleration by slow-mode eddies in compressible MHD turbulence is of
order~$l/v_{\rm A}$ and independent of~$p$ and particle species, aside
from a possible weak momentum and species dependence arising from the
$\ln(l_\parallel^\prime v_{\rm A}/D_\parallel)$~term.
The solution to equations~(\ref{eq:fp0a}) through (\ref{eq:bc3}) is 
(Kardashev 1962, Schlickeiser 2002)\footnote{
See also Parker (1957),  Ramaty (1979), 
Schlickeiser (1984), Ball et~al (1992), and Park \& Petrosian (1995).}
\[
f = \frac{A}{2}\left(\frac{p}{p_0}\right)^{-(3+\sqrt{9+4\chi})/2}
\left[1 + {\rm erf}\left(\frac{-\ln(p/p_0)+\tau\sqrt{9+4\chi}}{2\sqrt{\tau}}\right)\right]
\]
\begin{equation} 
+
\frac{A}{2}\left(\frac{p}{p_0}\right)^{(-3+\sqrt{9+4\chi})/2}
\left[1-{\rm erf}\left(\frac{\ln(p/p_0)+\tau\sqrt{9+4\chi}}{2\sqrt{\tau}}\right)\right],
\label{eq:fsolvep} 
\end{equation} 
where
\begin{equation}
\tau \equiv \frac{t}{\tau_{\rm acc}},
\label{eq:deftau} 
\end{equation} 
\begin{equation}
\chi = \frac{\tau_{\rm acc}}{\tau_{\rm esc}},
\label{eq:defchi} 
\end{equation} 
and
\begin{equation}
{\rm erf}(x) = \frac{2}{\sqrt{\pi}}\int_0^x e^{-y^2} dy
\label{eq:erf} 
\end{equation} 
is the error function.

\begin{figure}[h]
\vspace{11cm}
\includegraphics{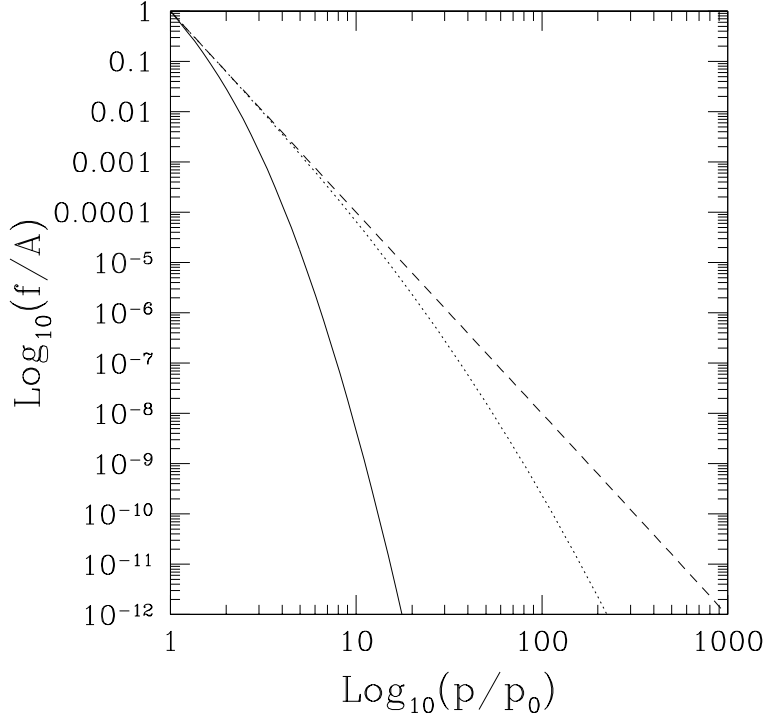}
\caption{The distribution function~$f$ at $t =  \tau_{\rm acc}/10$ (solid
line), $t= \tau_{\rm acc}/2$ (dotted line), and $t = 2 \tau_{\rm acc}$
(dashed line) for $\tau_{\rm acc} = 4\tau_{\rm esc}$, when
Coulomb losses are neglected.
\label{fig:f1}}
\end{figure}

The solution in equation~(\ref{eq:fsolvep}) is illustrated in
figure~\ref{fig:f1} for $\chi = 4$. The solid line corresponds
to~$\tau = 0.1$, the dotted line to~$\tau = 0.5$, and the dashed line
to~$\tau = 2$. Roughly speaking, $f$ approaches the power law $
A(p/p_0)^{-(3+\sqrt{9+4\chi})/2}$ for $p=p_0$ up to $p=p_{\rm max}
\sim p_0 e^{\tau\sqrt{9 + 4 \chi}}$, with $f$ dropping off sharply
for~$p>p_{\rm max}$. The width of the cutoff in $\ln(p/p_0)$-space
increases like~$\sqrt{\tau}$.  For the case~$\chi=4$ plotted in
figure~\ref{fig:f1}, $f\sim p^{-4}$ for $p \lesssim p_{\rm max} = p_0
e^{5 \tau}$.  When $\tau_{\rm acc} \gg \tau_{\rm esc}$, $p_{\rm max}
\sim e^{2t/\sqrt{\tau_{\rm acc} \tau_{\rm esc}}}$, and the de-facto
time scale for particle acceleration is~$\sim \sqrt{\tau_{\rm esc}
\tau_{\rm acc}}$.  This time scale is short for small~$\tau_{\rm esc}$
because the only particles to reach high energies when $\tau_{\rm esc}
\ll \tau_{\rm acc}$ are a small fraction that happen to be accelerated
much more rapidly than average.

The energy spectrum~$N(E)$ is the number of accelerated particles per
unit energy. In the non-relativistic limit, it is given by
\begin{equation}
N(E) = 4\pi m p f(p)\left| \begin{array}{l}
\hspace{0.3cm} \\
\hspace{-0.2cm} _{p= \sqrt{2 mE}}
\end{array}\right.,
\label{eq:defN} 
\end{equation} 
where $m$ is the particle mass.  In the non-relativistic limit, $N(E)$
approaches the power law $E^{-q}$ for $E< p_{\rm max}^2/2m$, with $q =
(1 + \sqrt{9 +4\chi})/4$; equivalently, a specified power law $N(E)
\propto E^{-q}$ corresponds to~$\chi = 4 q^2 - 2q -2$.  In the
ultra-relativistic limit, $N(E)$ approaches the power law~$E^{-q}$ for
$E< p_{\rm max}c$, with $q= (-1+ \sqrt{9+4\chi})/4$; equivalently, for
a specified~$q$, $\chi = 4q^2+ 2q -2$.  If escape is neglected,
then~$\chi = 0$; this corresponds to~$q=1$ in the non-relativistic limit and~$q=1/2$
in the ultra-relativistic limit.

\section{Steady-state energy spectrum of accelerated electrons
with Coulomb losses and an energy-independent escape time}
\label{sec:stst} 

\begin{figure}[h]
\vspace{11cm}
\includegraphics{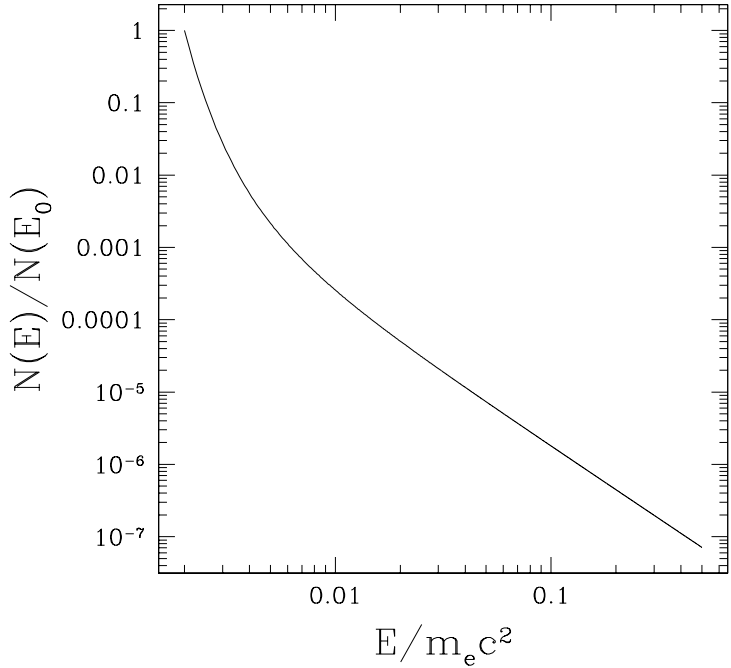}
\caption{The steady-state energy spectrum of accelerated electrons
taking into account Coulomb losses and an energy-independent escape
time, assuming $E_0 = 0.002 \mbox{ $m_e$c}^2$, $n=2 \times
10^{9}\mbox{ cm}^{-3}$, $\chi = 10$, and $\tau_{\rm acc} = 5$~s.
\label{fig:f2}}
\end{figure}

In the presence of Coulomb losses, steady electron injection
at energy~$E_0$,  and  an energy-independent
escape time~$\tau_{\rm esc}$, the energy spectrum of superthermal
electrons $N(E)$ satisfies the Fokker-Planck equation
(Park \& Petrosian 1995)
\begin{equation}
\frac{\partial N}{\partial t}
= \frac{\partial^2}{\partial E^2}\left( D_E N \right)
- \frac{\partial }{\partial E}\left( A_E N\right) - \frac{N}{\tau_{\rm esc}}
+ S \delta(E-E_0) \Theta(t),
\label{eq:fpn} 
\end{equation} 
where in the non-relativistic limit
\begin{equation}
D_E = \frac{p^2 D_p}{m_e^2},
\label{eq:defDE} 
\end{equation} 
\begin{equation}
A_E = \frac{m_e}{p^2} \frac{d}{dp}\left(p D_E\right) - \frac{\alpha_c  m_e^2c^3}{p},
\label{eq:defAE} 
\end{equation} 
\begin{equation} 
\alpha_c = 6\times 10^{-3} \left(\frac{n}{10^{10}\mbox{ cm}^{-3}}\right) \mbox{ s}^{-1},
\label{eq:defac} 
\end{equation} 
$n$ is the electron density,
$S$ is a constant, and~$\Theta(t)$ is the Heavyside function.
Assuming
\begin{equation}
D_p = \frac{p^2}{\tau_{\rm acc}},
\end{equation} 
with $\tau_{\rm acc}$ independent of~$p$,
the steady-state ($t\rightarrow \infty$) solution to equation~(\ref{eq:fpn}) 
for $E>E_0$ is [Park \& Petrosian 1995, equation (71)]\footnote{Equation~(\ref{eq:solN}) of
this paper follows from the results of Park \& Petrosian (1995) upon noting
two minor typos  in that paper. In particular, 
the second line of equation (67) of that paper should read 
$\delta_\pm = [(a +1)/2] \pm \mu$, while the second half
of equation (68) of that paper  should read $\lambda_0 = [(a+1)/2]^2 + \theta$
(Petrosian 2003).
The correct versions of the equations were given by Park (1995).}
\begin{equation}
N(E) = c_1 x^{3\tilde{\gamma}/2} M(\tilde{a}, \tilde{b}, |\beta|x^{-3/2}),
\label{eq:solN} 
\end{equation} 
where $c_1$ is a constant, 
\begin{equation}
x= \frac{E}{m_e c^2},
\label{eq:defx} 
\end{equation} 
\begin{equation}
\tilde{a} = \frac{1}{2} + \frac{\sqrt{9 + 4\chi}}{6},
\label{eq:at} 
\end{equation} 
\begin{equation}
\tilde{b} = 1 + \frac{\sqrt{9 + 4\chi}}{3},
\label{eq:bt} 
\end{equation} 
\begin{equation} 
\tilde{\gamma}=  - \frac{1}{6} - \frac{\sqrt{9 + 4\chi}}{6},
\label{eq:defg} 
\end{equation} 
and
\begin{equation}
\beta = \frac{\alpha_c \tau_{\rm acc}}{6\sqrt{2}}.
\label{eq:defbeta} 
\end{equation} 
Equation~(\ref{eq:solN}) is illustrated in figure~\ref{fig:f2} for 
the parameters $E_0 = 0.002 \mbox{ $m_e$c}^2$,
$n=2 \times 10^{9}\mbox{ cm}^{-3}$, $\chi = 10$, and
$\tau_{\rm acc} = 5$~s.
For $x  \gg \beta^{2/3}$, Coulomb losses become unimportant, and the
spectrum obtains the same steady-state power law as in section~\ref{sec:spectra}.
For $x\lesssim \beta^{2/3}$, Coulomb losses cause the spectrum to
steepen relative to its high-energy power~law (Hamilton \& Petrosian 1992).

\section{Application to electron acceleration in solar flares}
\label{sec:sf} 

In this section it is shown that the acceleration mechanism discussed
in this paper is probably not responsible for electron acceleration in
solar flares in the 10-100~keV range.  On the other hand, the
predicted energy spectrum is roughly consistent with the energy
spectra of electrons in the 100-5000~keV range, although it is not
clear that~$D_\parallel \ll l v_{\rm A}$ for such particles.

\begin{figure*}[h]
\vspace{11cm}
\includegraphics{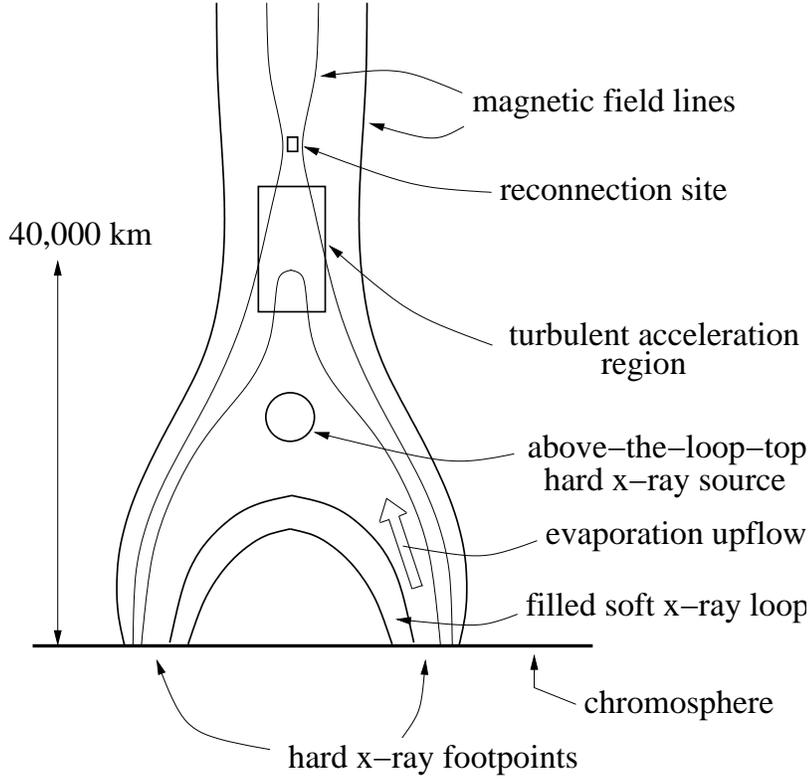}
\caption{Reconnection model for a solar flare with a single
pair of footpoints.
\label{fig:f3}}
\end{figure*}

A simple and fairly standard flare model is considered, as depicted in
figure~\ref{fig:f3}, with features drawn from
the works of, e.g., Sturrock (1966),  LaRosa
et~al (1994), Tsuneta (1996), 
Miller et~al (1996), Tsuneta et~al (1997), Miller et~al (1997), and
Aschwanden (2002). The flare
arises from reconnection between open magnetic field lines above a
pair of footpoints.  The reconnection layer is assumed to be
Petschek-like, so that the reconnection rate is sufficiently
fast. Most of the magnetic energy is released outside (downstream) of
the reconnection layer, where the magnetic tension associated with
newly reconnected field lines accelerates plasma away from the
reconnection site in a ``sling-shot'' action.  It is assumed that
turbulence is generated in this magnetized outflow with rms
velocity~$\sim v_{\rm A}$, and that the
turbulence is similar to strong homogeneous MHD turbulence
(section~\ref{sec:MHD}).  Electrons accelerated by the turbulence
stream towards the solar surface, evaporating chromospheric plasma and
emitting hard bremsstrahlung x-rays at the pair of footpoints that
anchor the magnetic flux tube occupied by the energetic electrons. The
evaporated chromospheric plasma rises, filling a magnetic-flux-tube
loop with hot relatively dense plasma that emits soft x-rays. As
additional field lines reconnect, the hard x-ray footpoints move
progressively outward and the top of the soft-x-ray loop rises.
Time-of-flight measurements typically place the electron acceleration
region well above the tops of the soft x-ray loops and also above the
hard x-ray emission that has been observed above the soft x-ray loops
in a number of flares (Aschwanden 2002).

The electron energy spectrum varies between flares and between the
footpoints and above-the-looptop regions. If $N(E)$ is taken to be a
power-law~$E^{-q}$ in the 10-100~keV range, then a typical value
of~$q$ is~4, although values between~$1.5$ and~$\sim 10$ have been
observed (Alexander \& Metcalf 1997, Miller et~al 1997, Petrosian
2002). Above~$\sim 100$~keV, the electron spectrum flattens, with
values of~$q$ in the range~$2.0-2.7$ for the three flares modeled by
Kundu et~al (2001). Time-of-flight measurements indicate that the
electron acceleration time is very short.  For the flare of
December~15, 1991, 1932 UT, Aschwanden (2002) found that the time to
energize electrons to~511~keV is~$\ll 0.36$~s; Aschwanden (2002)
pointed out that even the
$0.18$~s acceleration time to~511~keV found by Miller et~al (1996) is
too long to be consistent with the energy-dependent arrival times of electrons
at the hard-x-ray footpoints in this flare.

The start frequencies of bi-directional and type~III radio bursts
indicate that the electron density in the acceleration
region is in the range~$n_e = 0.6 - 10 \times 10^9 \mbox{ cm}^{-3}$
(Aschwanden \& Benz 1997). Models of the gyrosynchrotron emission
suggest that the magnetic field strength at the tops of flaring
loops is~$\sim 300$~G~(Kundu et~al 2001). The magnetic field in the acceleration
region  is probably somewhat smaller. For typical parameters, it
is assumed that $n_e = 2\times 10^9 \mbox{ cm}^{-3}$ and $B= 200$~G
in the acceleration region, yielding~$v_{\rm A} \sim 10^4$~km/s.
It is assumed that the outer scale
of the turbulence is $l=5000$~km, and that the region of 
greatest turbulence intensity is of size~$\sim l$.

It has been hypothesized that microturbulence in the acceleration region, 
e.g., whistler turbulence,
causes~$\lambda_{\rm mfp}$ for energetic electrons to be much shorter
than the Coulomb mean free path~(Miller et~al 1997).  If $D_\parallel
\ll l v_{\rm A}$ as assumed in this paper, then the time for particles
to diffuse out of the region of strongest turbulence, $l^2/D_\parallel$, is much
longer than the acceleration time,~$\tau_{\rm acc}\sim l/v_{\rm A}$.
On the other hand, the plasma undergoes a bulk flow at speed~$\sim
v_{\rm A}$ away from the reconnection site that advects electrons out
of the region of strongest turbulence on a time scale~$l/v_{\rm A}$.
For particles with~$D_\parallel \ll l v_{\rm A}$, transport out of the
turbulent region is dominated by this advection, and the escape
time~$\tau_{\rm esc}$ is then~$\sim l/v_{\rm A} \sim 0.5$~s, independent of
electron energy~(Blackman 2003).
If $d_{\rm min} v_{\rm A} \ll D_\parallel
\ll l v_{\rm A}$, then particle acceleration by slow-modes is
approximately described by equation~(\ref{eq:dp4a}). Taking
$l_\parallel^\prime \rightarrow l$ in equation~(\ref{eq:dp4a}) and
$D_\parallel = l v_{\rm A}/10$, one finds that $\tau_{\rm acc}=p^2/D_p
\sim 10l/ v_{\rm A} \sim 5$~s, and $\chi = \tau_{\rm acc}/\tau_{\rm esc} = 10$.

Coulomb losses are the dominant energy losses for electrons in solar
flares with $E< 100$~keV (Hamilton \& Petrosian 1992). If escape is
modeled with an ad-hoc loss term as in equation~(\ref{eq:fpn}), then
the energy spectrum of electrons accelerated by slow modes approaches
the form depicted in figure~\ref{fig:f2}. For energies above a
few~keV, Coulomb losses can be neglected and $N(E)\propto E^{-q}$,
with~$q=2$. Such a spectrum is significantly harder than typical
solar-flare spectra in the 10-100~keV range, as discussed above.  For energies above a few
keV, the time dependent spectrum behaves approximately as in
section~\ref{sec:spectra}, with $N(E) \propto E^{-2}$ for $E< E_{\rm
max}$, with $E_{\rm max} = p_{\rm max}^2/2m \propto e^{\rm
(2t/\tau_{\rm acc}) \sqrt{9 + 4\chi}} = e^{\rm 14 t/\tau_{\rm
acc}}$. The time required for $E_{\rm max}$ to increase from 5~keV to
500~keV is then~$\sim 0.3\tau_{\rm acc} \sim 1.5$~s.  This time is
more than a factor of 10 larger than the energization time implied by
time-of-flight measurements for the flare of December 15, 1991, 1932
UT (Aschwanden 2002). The inability of the acceleration mechanism described in this
paper to explain the observed steep electron spectra and short
acceleration times makes it an unlikely explanation for electron
acceleration in solar flares in the 10-100~keV range. 

The predicted spectra are roughly consistent with the energy spectra of electrons in the
100-5000~keV range inferred from microwave gyrosynchrotron emission (Kundu et~al 2001).
The electrons responsible for this emission may have considerably longer
acceleration times than the electrons involved in x-ray bursts.
The acceleration mechanism discussed in this paper could
explain these high-energy electrons, but only if $D_\parallel
\ll l v_{\rm A}$ for electrons in this energy range.

\section{Summary}
\label{sec:conc} 

The momentum diffusion coefficient~$D_p$ arising from slow modes in
strong compressible anisotropic MHD turbulence is calculated for the
case that the rms turbulent velocity is~$\sim v_{\rm A}$, pitch-angle
scattering of energetic particles is very efficient ($d_{\rm min}
v_{\rm A} \ll D_\parallel \ll l v_{\rm A}$), the energetic particle
speed~$v$ is~$\gg v_{\rm A}$, and $\beta \lesssim 1$. It is found that
\begin{equation} 
D_p \simeq \frac{2 p^2 v_{\rm A}}{9  l}\; \left[\ln 
\left(\frac{l v_{\rm A}}{D_\parallel}\right) + 2\gamma - 3 \right],
\mbox{ \hspace{0.3cm} for $v_{\rm A} d_{\rm min} \ll D_\parallel \ll v_{\rm A} l$,}
\label{eq:dpsolve2} 
\end{equation} 
where $\gamma \simeq 0.577$ is Euler's constant.  Aside from a
possible weak momentum dependence associated with the $\ln (l v_{\rm
A}/D_\parallel)$ term, $D_p\propto p^2$, implying that the
acceleration time~$\tau_{\rm acc} = p^2/D_p$ is independent of~$p$. In
addition, $\tau_{\rm acc}$ is of order~$l/v_{\rm A}$ and approximately
independent of particle species.

Slow modes in the $D_\parallel \ll l v_{\rm A}$-limit are an unlikely
explanation for electron acceleration in solar flares to energies of
10-100~keV, because for solar-flare conditions the predicted
acceleration times are too long and the predicted energy spectra are
too hard.  The acceleration mechanism discussed in this paper could in
principle explain the relatively hard spectra of
gyrosynchrotron-emitting electrons in the~100-5000~keV range, but only
if $D_\parallel \ll l v_{\rm A}$ for such particles.

\acknowledgements 

I thank Eliot Quataert, Steve Spangler, Torsten Ensslin, Eric
Blackman, Ted LaRosa, and Vahe Petrosian for valuable input.  I also thank the
referee, R. Schlickeiser, for his very helpful comments. This work was
supported by NSF grant AST-0098086 and DOE grants DE-FG02-01ER54658
and DE-FC02-01ER54651.

Alexander, D., \& Metcalf, T. 1997, ApJ, 489, 442

Aschwanden, M. 2002, Sp. Sci. Rev., 101, 1

Aschwanden, M., \& Benz, A. 1997, ApJ, 480, 825

Ball, L., Melrose, D., \& Norman, C. 1992, ApJ, 398, L65

Berezinskii, V., Bulanov, S., Dogiel, V., Ginzburg, V.,
\& Ptuskin, V. 1990, {\em Astrophysics of Cosmic Rays} (New York: Elsevier)

Bhattacharjee, A., \& Ng, C. S. 2000, ApJ, submitted

Blackman, E. 2003, private communication

Bykov, A., \& Toptygin, I. 1982, J. Geophys., 50, 221

Bykov, A., \& Toptygin, I., 1990, Sov. Phys. JETP, 71, 702

Bykov, A., \& Toptygin, I., 1993, Physics Uspekhi, 36, 1020

Chandran, B. 2000, Phys. Rev. Lett., 85, 4656

Chandran, B., \& Maron, J. 2003, ApJ, accepted

Cho, J., \& Lazarian, A. 2002, Phys. Rev. Lett., 88, 245001

Cho, J., \& Vishniac, E. 2000, ApJ, 539, 273

Dolginov, A., \& Silant'ev, N. 1990, A\&A, 236, 519

Ghosh, S., \& Goldstein, M. 1997, J. Plasma Phys., 57, 129

Sridhar, S., \& Goldreich, P. 1994, ApJ, 432, 612

Goldreich, P., \& Sridhar, S. 1995, ApJ, 438, 763

Goldreich, P., \& Sridhar, S. 1997, ApJ, 485, 680

Gruzinov, A., \& Quataert, E. 1999, ApJ, 520, 849

Hamilton, R., \& Petrosian, V. 1992, ApJ, 398, 350

Higdon, J. C. 1984, ApJ, 285, 109

Higdon, J. C. 1986, ApJ, 309, 342

Jokipii, J. R. 1966, ApJ, 146, 480

Kardashev, N. 1962, Sov. Astr., 6, 317

Katz, M., \& Stehlik, M. 1991, Astr. Sp. Sci., 183, 259

Kulsrud, R., \&  Ferrari, A. 1971, Astrophys. Sp. Sci., 12, 302

Kulsrud, R., \& Pearce, W. 1969, ApJ, 156, 445

Kundu, M., Nindos, A., White, S., \& Grechnev, V. 2001, ApJ, 557, 880

LaRosa, T., Moore, R., \& Shore, S. 1994, ApJ, 425, 856

Lithwick, Y., \& Goldreich, P. 2001, ApJ, 562, 279

Maron, J., \& Goldreich, P. 2001, ApJ, 554, 1175

Matthaeus, W., Oughton, S., \& Ghosh, S., Phys. Rev. Lett., 81, 2056

Milano, L., Matthaeus, W., Dmitruk, P., Montgomery, D. C. 2001,
Phys. Plasmas,  8, 2673

Miller, J. 1998, Sp. Sci. Rev., 86, 79

Miller, J., Cargill, P., Emslie, A., Homan, G., Dennis, B., LaRosa, T.,
Winglee, R., Benka, S., \& Tsuneta, S. 1997, J. Geophys. Res., 102, 14631

Miller, J.,  LaRosa, T., \& Moore, R. 1996, ApJ, 461, 445

Montgomery, D., \& Matthaeus, W. 1995, ApJ, 447, 706

Montgomery, D., \& Turner, L. 1981, Phys. Fluids, 24, 825

Oughton, S., Priest, E., \& Matthaeus, W. 1994, J. Fluid Mech., 280, 95

Park, B. 1995, Ph.D. thesis, Stanford University

Park, B., \& Petrosian, V. 1995, ApJ, 446, 699

Parker, E. 1957, Phys. Rev., 107, 830

Petrosian, V. 2002, astro-ph/0207482

Petrosian, V. 2003, private communication

Ptuskin, V. 1988, Sov. Astron. Lett., 14, 255

Quataert, E. 1998, ApJ, 500, 978

Ramaty, R. 1979, in {\em Particle Acceleration Mechanisms in Astrophysics},
ed. Arons, J., Max, C., McKee, C. (AIP: New York), p. 135

Schlickeiser, R. 1984, A\&A, 136, 227

Schlickeiser, R. 2002, {\em Cosmic Ray Astrophysics} (Berlin: Springer)

Schlickeiser, R., \&  Miller, J. 1998, ApJ, 492, 352

Shebalin, J. V., Matthaeus, W., \& Montgomery, D. 1983, J. Plasma Phys., 29, 525

Skilling, J. 1975, MNRAS, 172, 557

Spangler, S. 1999, ApJ, 522, 879

Stone, J., Ostriker, E., \& Gammie, C. 1998, ApJ, 508, L99

Sturrock, P. 1966, Nature, 211, 695

Tsuneta, S. 1996, ApJ, 456, 840

Tsuneta, S., Masuda, S., Kosugi, T., \& Sato, J. 1997, ApJ, 478, 787

Yan, H., \& Lazarian, A. 2002, Phys. Rev. Lett., 89, 281102

\end{document}